\documentclass[aps,reprint,superscriptaddress,amsmath,amssymb]{revtex4-2}

\usepackage{graphicx}
\usepackage{bm}
\usepackage{physics}
\usepackage{hyperref}
\usepackage{orcidlink}
\usepackage{mathrsfs}
\usepackage{array}
\usepackage{amsmath}
\usepackage{subfigure}
\usepackage{amstext}
\usepackage{amsfonts}
\usepackage{amssymb}
\usepackage{epstopdf}

\newcommand{\+}[0]{\dagger}

\newcommand{\x}[0]{{\cdot}}
\newcommand{\e}[0]{\bm{\epsilon}}
\newcommand{\ve}[0]{\bm{\varepsilon}}
\newcommand{\g}[0]{\textsc{g}}
\newcommand{\bc}[0]{\textsc{b}}

\newcommand{\sys}[0]{\textsc{s}}
\newcommand{\env}[0]{\textsc{e}}
\newcommand{\tot}[0]{\mathrm{tot}}
\newcommand{\mat}[1]{\begin{pmatrix} #1 \end{pmatrix}}

\begin{document}

\title{From Reversible Quantum Dynamics to Statistical Probability:\\
A dynamical solution to  the origin of probability and Hilbert's sixth problem}

\author{Wei-Min Zhang\orcidlink{0000-0003-2117-3608}}
\email{wzhang@mail.ncku.edu.tw}
\affiliation{Department of Physics, National Cheng Kung University, Tainan 70101}
\date{Sept 16, 2026, Updated Sept. 28, 2026}

\begin{abstract}
Hilbert's sixth problem placed probability and mechanics at the center of the axiomatization of physics. 
In modern form, it asks how irreversible statistical probability and 
thermodynamics can follow from the reversible, deterministic dynamics of a closed  
system. Building on the exact open quantum system theory we developed over the past two decades, 
we construct a dynamical route from the reversible quantum dynamics to the reduced statistics. 
Assuming the system and its environment form a closed ``universe", for bosonic 
or fermionic system linearly coupled to arbitrary environments, we can derive the
exact reduced density operator determined fully by a dissipative propagator $\bm u$ and a fluctuation 
correlation function $\bm v$, where the generalized fluctuation--dissipation theorem naturally holds. 
For continuous spectral density with no bound-state pole, 
$\bm u$ vanishes in the long-time limit, information about the system initial state is lost completely, and the 
reduced density operator approaches a thermal Gibbs state of a non-perturbatively renormalized 
Hamiltonian. Strong system-environment coupling
and gapped spectral density with localized bound states instead preserve long-time memory and prevent thermalization. 
We then remove the usual assumption of an initially thermal, mixed environment, 
start from a product of pure states for the system and environment with every environmental mode in squeezing vacuum, unitary evolution 
dynamically generates system-environment entanglement and a periodic quasi-Gibbs reduced state for a finite environment. 
When the number of environmental degrees of freedom tends to infinity, the spectrum becomes continuous, 
coherence dephases, the recurrence disappears, and the exact reduced 
density operator converges to a thermal Gibbs state $\rho_\g=Z_\sys^{-1}\exp\big(\!-\!\beta_{\rm eff} \widetilde{H}_\sys\big)$. 
The total entropy of the system plus environment remains zero, whereas the system entropy reaches its  
maximum. Probability is therefore not an additional random postulate imposed on unitary/reversible evolution, 
it is an intrinsic dynamical outcome arising from the subsystem dynamics within a closed composite system. 
These results provide an exact, testable dynamical solution, elucidating how deterministic reversible evolution 
gives rise to statistical probabilities.
\end{abstract}


\maketitle

\section{Introduction: The Modern Physical Core of Hilbert's Sixth Problem}

In 1900, Hilbert called for an axiomatic treatment of those physical sciences in which mathematics 
already played an important part, placing the theory of probability and mechanics first. He singled out the problem 
with two core issues: ``(i) the axiomatic treatment of probability with limit theorems for the foundation of statistical physics;
(ii) the rigorous theory of limiting processes which lead from the atomistic view to the laws of motion of continua.''
These two foundational issues are generally known as the main part of Hilbert's sixth problem~\cite{Hilbert1902}. 
Hilbert's sixth problem arose at the interface of classical mechanics, kinetic theory, and continuum physics, 
but its physical content should not be frozen at the scientific frontier of 1900. After 1900, physics sciences 
have made tremendous achievements because of the discoveries of relativistic theory and quantum mechanics, 
while Newtonian mechanics has been found to be a non-relativistic limit and a classical limit, respectively.

In fact, the first core issue in Hilbert's sixth problem about the axiomatic treatment of probability was soon solved 
mathematically in 1933  by Kolmogorov who established the foundation of the theory of probability in terms of  measure 
theory  \cite{Kolmogorov1956}. At the same time, this issue was also formulated at quantum level
in 1932 by von Neumann for the axiomatic quantum mechanics  \cite{Neumann1955}.  
From 1936 to 1943, collaborating with Murray, von Neumann further extended 
his axiomatic quantum mechanics to more general algebraic structures, known as von Neumann algebras 
or non-commutative measure theory
\cite{MurrayNeumann1937a,MurrayNeumann1937b,MurrayNeumann1940,MurrayNeumann1943}.
Kolmogorov's theory became a classical limit of non-commutative measure theory. 
In 1970's, von Neumann algebras were considered as a basis of axiomatic quantum field 
theory \cite{Haag1992}.  

After quantum mechanics, quantum field theory, and gauge theories were established, the second core 
issue in Hilbert's sixth problem acquired a more foundational formulation: 
Can statistical probability be produced by microscopic unitary quantum evolution itself, without introducing
wavefunction probability interpretation, inserting a probability ensemble or a thermal bath at beginning 
so that one can go far beyond Boltzmann's original investigation of statistical mechanics based on 
the principle of Newtonian mechanics \cite{Boltzmann1872}.
The foundational equations of classical and quantum mechanics are deterministic. For a closed quantum 
system, Schr\"odinger equation and equivalently von Neumann equations,
\begin{equation}
 i\hbar\frac{d}{dt}\ket{\Psi(t)}=H_{\rm tot}\ket{\Psi(t)},~~
 \frac{d}{dt}\rho_\tot(t)=\frac{1}{i\hbar}[H_\tot,\rho_\tot(t)],
\end{equation}
preserve purity, total entropy, and information under unitary (reversible) evolution. This appears to conflict with 
irreversible equilibration physics. 

Statistical physics was built on the equilibrium postulate that after 
a sufficiently long time, a given large-scale system containing $\sim 10^{23}$ particles and very weakly interacting with its 
environment can always reach a thermal equilibrium, and the equilibrium statistical probability distribution $\rho$ does 
not depend on its initial state (thermal irreversibility). This thermalization process is the foundation of statistical 
mechanics that has remained unsolved for the past century and a half~\cite{Huang1987}. More specifically,
it requires to answer two time-honored questions: (i) how does macroscopic irreversibility emerge from microscopic 
reversibility; (ii) how does the system relax to thermal equilibrium with its environment?
Many physicists, including the early works by Gibbs \cite{Gibbs1902} and by Einstein~\cite{Einstein1902}, 
attempted to solve this thermalization problem 
after Boltzmann's original work in 1872 \cite{Boltzmann1872}.  
In that work~\cite{Boltzmann1872}, Boltzmann derived Boltzmann equation 
through the particle-particle collisions with the assumption of molecular chaos (stosszahlansatz) and some other 
approximations for ideal, dilute gas and introduced the quantitative H-theorem showing how molecular collisions 
drive a gas toward equilibrium. However, Boltzmann's work was challenged by the 
Loschmidt’s reversibility paradox \cite{Loschmidt1876} and by Zermelo’s recurrence paradox 
\cite{Zermelo1896} based on deterministic Newtonian mechanics.

On the other hand, statistical mechanics based on the equilibrium postulate can derive the canonical or 
grand-canonical Gibbs state through entropy maximization~\cite{Kolmogorov1956,Neumann1955,Jaynes1957,Callen1985},
\begin{equation}
 S[\rho]=-k_B\Tr(\rho\ln\rho) \longrightarrow
 \rho_\g=Z^{-1}e^{-\beta(H-\mu N)},
\label{eq:Gibbs}
\end{equation}
The equilibrium postulate with Gibbs state $\rho_\g$ is extraordinarily successful in describing equilibrium 
statistical physics, modern material sciences and phase transitions, including quantum matters and 
quantum phase transitions which are currently still very active. 
But the postulate does not by itself derive the probability distribution from reversible microscopic motion of particles.

In fact, a pure state in Schr\"odinger picture nevertheless remains pure as long as the object is treated 
as closed. Statistical structure of quantum mechanics must therefore arise in the change of descriptive 
level to the subsystem (the measured system) from a composite that includes the measuring apparatus 
as a whole \cite{Neumann1955VI}. 
More general, combining a system $\sys$ and 
environment $\env$ into one closed composite (called as a closed ``universe") while tracing over exactly all environment 
states gives the reduced density operator 
of the system
\begin{equation}
 \rho_\sys(t)=\Tr_\env\!\left[U(t,t_0)\rho_\tot(t_0)U^\dagger(t,t_0)  \right],
\label{eq:reduced}
\end{equation}
where $\rho_{\rm tot}(t_0)$ represents the initial state of the closed ``universe". 
We found very recently that when traced exactly over all environment states, 
the system dynamics manifests ``internal causality breaking" so that the reduced 
density matrix $\rho_\sys(t)$ becomes a mixture even if $\rho_\tot(t)$ remains pure \cite{Yang2025}. 
Probability is then not an interpretation to the total dynamics. It is the spectrum of statistical weights carried 
by the reduced density operator of the system after inaccessible environmental 
degrees of freedom have been traced out or a measurement on the system has been performed in quantum mechanics.

The Feynman--Vernon influence functional expresses this reduction as an exact path integral over environmental 
classical paths linearly coupling Brownian particle and reveals the origin of dissipation and fluctuations~\cite{Feynman1963}.
This influence functional has been subsequently exposed further the physics of dissipation and fluctuations in 
system-environment back-actions~\cite{Mohring1980,Caldeira1983,Hu1992}. 
The framework used in this paper formulates in a different way. We take the trace as an integral over the 
coherent-state complex space and Grassmann space for boson and fermion systems, respectively \cite{Zhang1990}, 
from which, with or without using the influence functional, we derived the exact equation of motion (master 
equation) of the reduced density operator $\rho_\sys(t)$ for a large class of quadratic Hamiltonians \cite{Tu2008,Jin2010,Lei2012,Zhang2012,Zhang2019}. 
The resulting reduced density operator $\rho_\sys(t)$ is fully determined by the dissipative and fluctuation 
correlation Green's functions where the generalized non-equilibrium fluctuation-dissipation theorem
naturally holds~\cite{Tu2008,Jin2010,Lei2012,Zhang2012,Zhang2019}.
The reduced density operator  
$\rho_\sys(t)$ and the master equation encapsulate all information of the open system (objects) that provide indeed the laws of 
motion of continua in Hilbert's sixth problem. The non-equilibrium fluctuation correlation function 
is the exact solution of the quantum Boltzmann equation in linear quantum systems. Fluid transport dynamics is also 
rigorously governed by the master equation.  
This provides a rigorous theory showing how irreversibility and statistical probability emerge from the unitary 
quantum evolution of a closed composite ``universe".

This paper has three aims. First,  from a modern physics perspective, we recast the dynamical content of the 
second foundational issue in Hilbert's sixth problem in the language of quantum mechanics, as we have discussed 
above. Second, we show in a self-consistent and self-sufficient manner the exact master equation of open quantum 
systems, and the solution of the single-particle distribution and the solution of the transient transport current, 
as well as the continuous spectral conditions for thermalization.
Third, the new theoretical result in this paper, we derive both mixed-state generation in 
a finite pure composite and a genuine Gibbs limit when a pure system is enlarged to infinitely many degrees of freedom.
This enlarging process shows how  ``time arrow" emerges from the reversible quantum dynamics evolution, thus eliminating 
the challenges posted by the Loschmidt’s reversibility paradox and Zermelo’s recurrence paradox in original Boltzmann's work
based on Newtonian mechanics. Consequently, the origin of probability is revealed and the foundation of statistical 
mechanics is established. 

\section{Exact Open Quantum System Framework}

Consider a multimode (or multilevel) system bilinearly coupled to the environment comprising multiple reservoirs
as a closed composite, each reservoir in the environment possesses an infinite number of energy modes or 
an infinite number of degrees of freedom.
A general quadratic bosonic or fermionic  composite with particle exchange and pairing excitations may 
be written as a Bogoliubov-de~Gennes Hamiltonian form,
\begin{align}
 H_\tot={}H_\sys + & H_\env  +H_{\sys\env} = \bm{a}^\+ \x \ve_\sys \x \bm{a} + \sum_\alpha 
 \bm{b}_\alpha^\+ \x \e^\alpha_\env \x \bm{b}_\alpha \notag \\
&  + \sum_\alpha \mat{\bm{a}^\+ & \bm{a}} \!\! \mat{\bm{V}^\alpha_{\sys\env} & \bm{\Delta}^\alpha_{\sys\env}  \\ 
    \pm \bm{\Delta}^{\alpha *}_{\sys\env} & \pm \bm{V}^{\alpha *}_{\sys\env} } \!\! \mat{\bm{b}_\alpha \\ \bm{b}^\+_\alpha}.
\label{eq:Htot}
\end{align}
Here 
$\bm a^\dag = (a^\dag_1, a^\dag_2, \cdots, a^\dag_{N_S})$, $\bm b^\dag_\alpha = (b^\dag_{\alpha 1}, b^\dag_{\alpha 2}, 
b^\dag_{\alpha 3} \cdots) $, and $a^\dag_i$ and $b^\dag_{\alpha k}$ create excitations of the system and the reservoir $\alpha$ 
in the environment, respectively. The corresponding energy
spectrum $\ve_\sys$ and $\e^\alpha_\env$ are given by $N_\sys \! \times \! N_\sys$ and 
$N^\alpha_\env \! \times \! N^\alpha_\env$ diagonal matrices where $N^\alpha_\env \to \infty$. 
Bosons obey commutation relations and fermions anticommutation relations. 
This Hamiltonian can describe Fano resonance, photon localization, and dynamics of cavities in photonic crystals
(
without pairings, $\bm\Delta^\alpha_{\sys\env}=0$)~\cite{Lei2012,Xoing2010,Tan2011,Lo2015,Yang2018}; also electrons in 
solid states, quantum-dot and nanodevice transport (
also no pairings)
~\cite{Tu2008,Jin2010,Yang2014,Yang2017}; as well as
superconducting Bogoliubov modes, topological boundary states, squeezing dynamics in integrated
quantum photonics (contain pairings)~\cite{Lai2018,HuangTopological2020,HuangBrownian2022}, etc.
The total Hamiltonian generates strictly unitary (reversible) evolution of the closed 
composite, non-unitarity can be rigorously derived for the system.

For an initially factorized state,
\begin{equation}
 \rho_{\rm tot}(t_0)=\rho_\sys(t_0)\otimes\rho_\env(t_0),
\end{equation}
where the system can be arbitrary initial state, and the environment is often assumed in thermal states.
By the {\it exact} partial trace over all the environment states either using influence functional on complex/Grassmann 
space~\cite{Tu2008,Jin2010,Lei2012,Zhang2012,Zhang2019} or directly taking the {\it exact} partial trace  
over the von Neumann equation \cite{HuangTopological2020,HuangBrownian2022}, and then further solve 
{\it exactly} the reduced system dynamics with a fine-graining, 
we arrived the exact master equation of the reduced density operator
that is time-local but maintains all non-Markovian quantum memories,
\begin{align}
 \frac{d}{dt} \rho_\sys  (t) = & \frac{1}{i\hbar}[\widetilde H_\sys(t),\rho_\sys(t)] +\sum_{ij}  \Big\{\bm \gamma_{ij}(t)[2a_j \rho_\sys(t)a^\dag_i  \notag \\
 & - a^\dag_i a_j \rho_\sys(t) - \rho_\sys(t) a^\dag_i a_j]  +\widetilde{\bm \gamma}_{ij}(t)\mathcal [a^\dag_i \rho_\sys(t)a_j \notag \\
 &\pm a_j \rho_\sys(t)a^\dag_i \mp a^\dag_i a_j \rho_\sys(t) - 
  \rho_\sys(t) a_j a^\dag_i  ]\Big\}.
\label{eq:master}
\end{align}
This exact derivation shows that a subsystem inside a composite is already governed by an non-unitary equation of motion ~\cite{explain1}. 
Without loss of generality, we have letted $\bm{\Delta}^\alpha_{\sys\env}=0$ in the derivation, 
the exact master equation with pairings 
is presented in our early publications~\cite{HuangTopological2020,HuangBrownian2022}. 

The renormalized system Hamiltonian in Eq.~(\ref{eq:master}) is $\widetilde H_\sys(t)
=\bm a^\dagger\widetilde{\bm\ve}_\sys(t)\bm a$.  
The non-Markovianness is embedded in the dissipation/loss structure (the $\bm \gamma$-term) 
and fluctuation-induced gain-loss balance structure (the $\widetilde{\bm \gamma}$-term) 
that are derived non-perturbatively in our exact calculation of $\rho_\sys(t)$ \cite{Tu2008,Jin2010,Lei2012}.
These coefficients in Eq.~(\ref{eq:master}) are fully determined by two Green's functions, $\bm u$ and $\bm v$ (both are 
$N_\sys \!\times \! N_\sys$ matrices),
\begin{align}
 & \widetilde{\bm\ve}_\sys(t) =  \frac{i}{2}  \big[\dot {\bm u} {\bm u}^{-1} - {\rm H.c.}\big], 
 ~~~\bm\gamma(t)=-\frac12\left[\dot {\bm u} {\bm u}^{-1}+{\rm H.c.}\right],\nonumber\\
& \widetilde{\bm\gamma}(t)=\dot {\bm v}-\left[\dot {\bm u} {\bm u}^{-1}{\bm v}+{\rm H.c.}\right].
\label{eq:coefficients}
\end{align}
The dissipative propagator $\bm u = \bm u(t,t_0)$ is independent of the initial state,
 and carries the memories about the energy and information exchanges between the system and environment, 
 which are eventually dissipated into the continuum spectrum of the environment, 
whereas $\bm v = \bm v(t,t)$ contains particle distributions and correlations in the system
injected by environmental fluctuations so that it carries memory of the environment (reservoir) initial state. 

These two Green's functions obey the following time-convolutional integro-differential equations of motion,
\begin{subequations}
\begin{align}
& \frac{d}{d\tau} {\bm u}(\tau,t_0)\!+\!\frac{i}{\hbar}\bm\ve_\sys \bm u(\tau,t_0)
 \!+\!  \sum_\alpha \!\! \int_{t_0}^{\tau}  \!\!\!\! d\tau' \bm g^\alpha (\tau,\tau')\bm u(\tau',t_0)=0,
 \label{eq:u}\\
& \frac{d}{d\tau} {\bm v}(\tau,t)\!+\!\frac{i}{\hbar}\bm\ve_\sys \bm v(\tau,t)
 \!+\!  \sum_\alpha \!\! \int_{t_0}^{\tau} \!\!\!\! d\tau'\bm g^\alpha (\tau,\tau')\bm v(\tau',t) \notag \\
   &\qquad \qquad \qquad \qquad = \sum_\alpha \!\!  \int_{t_0}^{t} \!\!\! d\tau'\, \widetilde{\bm g}^\alpha(\tau,\tau')\bm u^\dag (\tau',t_0).
 \label{eq:v}
\end{align}
\end{subequations}
where $t_0\!\le\!\tau\!\le\!t$. The dissipatiion and fluctuation kernels are given by the system-reservoir couplings, the reservoir spectrum 
and the initial reservoir states as follows, 
\begin{subequations}
\begin{align}
 \bm{g}^\alpha_{ij}(\tau,\tau')& = \frac{1}{\hbar^2} \!\! \int \!\! \frac{d\epsilon}{2\pi}\bm J^\alpha_{ij}(\epsilon)e^{-\frac{i}{\hbar}\epsilon(\tau-\tau')}, \\
 \widetilde{\bm g}^\alpha_{ij}(\tau,\tau')&= \frac{1}{\hbar^2} \!\! \int \!\! \frac{d\epsilon}{2\pi} \bm J^\alpha_{ij}(\epsilon)
 n_\alpha({\epsilon},T^\alpha_0)e^{-\frac{i}{\hbar}\epsilon(\tau-\tau')} ,
\label{eq:kernels}
\end{align}
\end{subequations}
where $\bm J^\alpha_{ij}(\epsilon)\!=\!2\pi\sum_kV^\alpha_{ik}V_{jk}^{\alpha *}\delta(\epsilon\!-\!\epsilon_k)$ is the system-reservoir-$\alpha$
spectral density matrix, 
and $n_\alpha({\epsilon},T^\alpha_0)$ is the corresponding initial particle distribution at the initial temperature 
$T^\alpha_0$ if the reservoir is initially in a thermal state. Different reservoir can have a different initial temperature.

It is particularly interesting to see that the inhomogeneous term in r.h.s of Eq.~(\ref{eq:v}) breaks 
the causality of the particle motion in the subsystem after the environmental degrees of freedom are exactly integrated out.
It tells explicitly that the correlation function $\bm v(\tau,t)$ at intermediate time $\tau$ is determined not only by the 
past dynamics from $t_0$ to $\tau$ but also by the further dynamics from $\tau$ to $t$, even though
the closed composite obeys a complete causal evolution and this 
internal causality violation in the reduced dynamics cannot be seen 
when the reduced density operator is actually determined or measured at time $t$. 
The essence of internal causality breaking arises from non-local time correlations between the 
forward and backward propagators of the reduced dynamics when the environmental 
degrees of freedom are exactly integrated out~\cite{Yang2025}. This gives the source of how the original reversible deterministic 
quantum evolution of a closed composite makes irreversible dynamics emergence in the reduced  
system. This irreversible source is not caused solely by an initial environment thermal state, any initial pure  
state with non-classicality can also generate it, as we will see in the next section.  

In fact, the exact solution of Eq.~(\ref{eq:v})  is \cite{Jin2010,Lei2012,Zhang2012,Zhang2019} 
\begin{align}
\bm v(\tau,t)= \sum_\alpha \!\! \int_{t_0}^{\tau}\!\!dt_1\!\! \int_{t_0}^{t}\!\!dt_2\,
\bm u(\tau,t_1)\widetilde{\bm g}_\alpha(t_1,t_2)\bm u^\dagger(t,t_2).   \label{fdt}
\end{align}
This identity naturally gives the generalized non-equilibrium fluctuation-dissipation theorem~\cite{Zhang2012,Zhang2019}. 
In the equilibrium limit, it recovers the general Callen-Welton fluctuation-dissipation theorem or the Green-Kubo 
fluctuation-dissipation relation~\cite{Callen1951,Kubo1966}, and in
the high temperature limit it reproduces the Einstein's fluctuation-dissipation theorem \cite{Einstein1905}.
It is a consequence of the unitarity of the closed composite leading to irreversibility of its subsystem dynamics. 
It is also a foundational condition for any open quantum system being able to approach microscopical thermal equilibrium. 

To see explicitly, we can directly 
calculate the single-particle density matrix ${\bm f}_{ij}(t) \equiv \Tr_\sys[a^\dag_j a_i \rho_\sys(t)]$ from the master 
equation, resulting in a rigorous form of the quantum Boltzmann equation in linear systems \cite{Jin2010,HuangTopological2020}:
\begin{align}
\dot{\bm f} -\dot{\bm v}= \dot {\bm u} {\bm u}^{-1}(\bm f -\bm v)+(\bm f -\bm v) ({\bm u}^\dag)^{-1} \dot{\bm u}^\dag. \label{qBe}
\end{align}
The diagonal matrix element ${\bm f}_{ii}$ is the single particle distribution (occupation number) in system's energy level $i$, 
and the off-diagonal matrix element ${\bm f}_{i\neq j}$ represents the correlation (transition) between different levels $i$ and $j$. 
It shows that the solution of single-particle density matrix $\bm f$ is just $\bm v$, apart from an initial 
function~\cite{Jin2010,Lei2012,HuangTopological2020}:
\begin{align}
{\bm f}(t) = {\bm v}(t,t) + {\bm u}(t,t_0)\bm f(t_0) {\bm u}^\dag(t,t_0). \label{qBes}
\end{align}
In other words, the solution of quantum Boltzmann equation strictly fellows the non-equilibrium 
fluctuation-dissipation theorem via the relation of  fluctuation correlation $\bm v$ and dissipative
propagator $\bm u$ given by Eq.~(\ref{fdt}). 

Furthermore, as a manifestation of hydrodynamics, the transport current flowing from reservoir 
$\alpha$ into the system can be also calculated from the exact master equation. Rewriting the master 
equation Eq.~(\ref{eq:master}) in the form~\cite{Jin2010,Lei2012,HuangTopological2020,Yang2017}
\begin{align}
& {d\rho_\sys(t)\over dt} = \frac{1}{i\hbar}\big[H_\sys,\rho_\sys(t)\big]+ 
\! \sum_\alpha \! \big(\mathcal{L}^{+}_\alpha[\rho_\sys(t)] +\mathcal{L}^{-}_\alpha[\rho_\sys(t)]) ,   
\end{align}
where the current superoperators $\mathcal{L}^{+}[\rho_\sys(t)]$
and $\mathcal{L}^{-}[\rho_\sys(t)]$ (see the explicit form given in \cite{Jin2010,Lei2012,Yang2017,HuangTopological2020})
describe the particles transiting into the system and transiting out from the system, respectively.
Then the transient transport current flowing from the reservoir $\alpha$ into the system is given by  
\begin{align}
& I_\alpha(t) \equiv \!- \! \left\langle \! \frac{d \hat{N}_\alpha(t)}{dt} \! \right\rangle = \rm{Tr}_\sys \big[\mathcal{L}^{+}_\alpha[\rho(t)]\big] = 
\!-\! \rm{Tr}_\sys \big[\mathcal{L}^{-}_\alpha[\rho(t)]\big]  \notag \\
  & \qquad = - \frac{2}{\hbar}{\rm Re}\!\! \int^t_{t_0} \!\! d\tau \Tr_\sys \! \big[ \bm g_\alpha(t.\tau) \bm f(\tau,t) 
  	- \widetilde{\bm g}_\alpha (t,\tau) \bm u^\dag(t, \tau)].   \label{transpcurrent}
\end{align}
Here $\hat{N}_\alpha \!=\! \sum_k b^\dag_{\alpha k} b_{\alpha k}$ is the total particle number operator of reservoir $\alpha$, and
$\bm f(\tau, t) \!=\!\bm v(\tau,t) \!-\! {\bm u}(\tau,t_0)\bm f(t_0) {\bm u}^\dag(t,t_0)$ is an two-point correlation extension of Eq.~(\ref{qBes}).
It shows that fluid dynamics is also fully determined by the dissipative propagator $\bm u$ and fluctuation correlation $\bm v$.

The analytic structure of $\bm u(t,t_0)$ gives the central dynamical criterion. Without loss of generality, consider the 
case where the environment contains one reservoir. If $\bm J(\epsilon)$ is a continuum spectral density matrix
and Eq.~(\ref{eq:u}) has no isolated real-axis bound-state pole which requires that the system-environment coupling is
not larger than a critical coupling \cite{Zhang2012,Xiong2015}, then the continuum spectrum causes decoherence and decays:
\begin{subequations}
\begin{align}
& \lim_{t\to\infty} \! \bm u(t,t_0) = \lim_{t\to\infty} \! \int \!\! \frac{d\epsilon}{2\pi}  \boldsymbol{\mathcal{D}}(\epsilon) e^{-i\epsilon(t-t_0)} =0 , \notag \\
& \lim_{t\to\infty}\!\bm v(t,t)=\lim_{t\to\infty}\! \bm f (t) = \!\! \int \!\! \frac{d\epsilon}{2\pi} \boldsymbol{\mathcal{D}}(\epsilon)n(\epsilon,T_0) \equiv \overline{\bm n},
\label{eq:u0}
\end{align}
\end{subequations}
where $\boldsymbol{\mathcal{D}}(\epsilon)$ is the level-broadening spectrum of the system arisen from the coupling to the environment,
$\overline{\bm n}$ denotes the steady-state particle distributions in the spectrum $\boldsymbol{\mathcal{D}}(\epsilon)$
that manifests the equilibrium fluctuation-dissipation theorem.  Boltzmann attempted to derive such irreversibility through particle-particle collisions 
in Newtonian mechanics in 1872 without fully successful, but here it is simply a natural consequence of linear open systems within the unitary 
evolution of a closed composite \cite{HuangRenorm2022,Xiong2020}. 

With the above steady-state solution of  $\bm u$ and $\bm v$, the system initial state 
dependence disappears. 
The long-time steady state $\rho_\sys$ approaches to the renormalized Gibbs state 
\cite{Xiong2015,HuangRenorm2022,Xiong2020},
\begin{equation}
 \lim_{t\to\infty}\rho_\sys(t) 
 =Z_\sys^{-1}\exp\!\left\{-\beta_{\rm eff}
 [\widetilde H_\sys-\mu_{\rm eff}N_\sys]\right\}.
\label{eq:thermal}
\end{equation}
where the thermalized temperature $k_B T_{\rm eff}=1/\beta_{\rm eff}$ can be different from the initial reservoir 
temperature $T_0$. Only in the very weak coupling limit, $T_{\rm eff}=T_0$ \cite{HuangRenorm2022,Xiong2015}.
This is a dynamical solution of the thermalization process showing how the reversible quantum evolution of a 
closed ``universe" produces  statistical probability description of open quantum systems that many physicists 
dreamed to prove during the last century and a half. We also find that strong system-environment couplings or
gapped spectral densities $\bm J(\epsilon)$ can instead generate localized bound states \cite{Zhang2012}, leaving $\bm u(t,t_0)$ 
non-decaying terms of the form $\sum_\ell \bm Z_\ell e^{-i\omega_\ell(t-t_0)}$,  exact thermalization  
is then not reachable~\cite{Xiong2015,HuangRenorm2022,Xiong2020}. This establishes a dynamical realization of 
thermalization to localization transition.

\section{From Pure Initial States to Gibbs Statistics}

However, an initial thermal environment already contains a statistical ensemble in the beginning. 
To address the origin of probability, the initial thermal state assumption on environment must be removed.
We will solve this problem from pure deterministic initial states step by step with three cases: 
(i) $N_\sys=1, N_\env$ is finite; (ii) $N_\sys=1, N_\env \to \infty$ (continuum environment); 
(iii) $N_\sys \to \infty, N_\env \to \infty$ (both system and environment become continuum), 
from which we show the dynamical emergence of probability in the unitary quantum dynamics 
of a larger system and the emergence of time arrow in macroscopic dynamics.

Case (i): $N_\sys=1, N_\env$ is finite, which is a minimal pure-state model showing dynamical production of mixed state.
The mode Hamiltonian takes a simple form,
\begin{equation}
H_{\rm tot}=\hbar \omega_\sys a^\dagger a+ \!
\sum_k\hbar \omega_k b_k^\dagger b_k+ \!
 \sum_k \hbar(V_{k}a^\dagger b_k+V_{k}^*b_k^\dagger a),
\label{eq:finitemode}
\end{equation}
with a completely separable pure state
\begin{equation}
 \ket{\Psi_{\rm tot}(t_0)}
 =\ket{\alpha_\sys}\otimes\prod_k\ket{s_k},
 \qquad s_k=r_ke^{i\theta_k},   \label{inisq}
\end{equation}
where the system mode  is a Glauber coherent state $ \ket{\alpha_\sys}=\exp\!\big(\alpha_\sys a^\dag-\alpha_\sys^\ast a\big) \!\ket{0}$ 
and each environmental mode
in a squeezed vacuum state $|s_k \rangle=\exp \!\big(\frac{1}{2}(s_k {b_k^\dag}^2\!-\!s_k^\ast b_k^2)\big)\!\ket{0}$
which is pure but has deterministic normal and anomalous covariances,
\begin{equation}
 n_k=\sinh^2r_k,\qquad
 m_k=-\frac{1}{2}e^{i\theta_k}\sinh 2r_k.   \label{nanm}
\end{equation}
Thus, no thermal ensemble or classical random variable is present initially.  This is actually a straightforward extension of our recent work 
for a simple two-mode coupling system \cite{Yang2025}.

Exactly integrating out all the environmental mode (which is indeed an exact non-perturbative renormalization procedure)
gives an analytic solution of the reduced density operator $\rho_\sys(t)$ for system 
mode~\cite{Yang2025}. The result is a displaced--squeezed quasi-Gibbs state,
\begin{align}
 \rho_\sys(t) & ={\cal N}(t)e^{A^\dagger(t)}
 \Big[\sum_{n=0}^{\infty}\delta^n(t)\ket n\bra n\Big]e^{A(t)} .
\label{eq:squezquasi_g}
\end{align}
A quasi-Gibbs state means that $\delta^n(t)$ is a time-dependent probability weight that 
differs from the stationary thermal limit.
In this analytical solution, 
$A^\dag(t) = [(1\!-\!\delta(t))\alpha(t) \!+\! \beta(t)\alpha^*(t)]a^\dag\!+\!\frac{1}{2}\beta(t)a^{\dag 2}$ is a 
displacement and squeezing operator and 
\begin{subequations}
\label{eq:delta}
\begin{align}
& \alpha(t)  =u(t,t_0)\alpha_\sys , \\ 
&\beta(t)= \frac{v_{2}(t,t)}{(1+v_{1}(t,t))^2-|v_{2}(t,t)|^2} , \\
& \delta(t) =1 -\frac{1+v_{1}(t,t)}{(1+v_{1}(t,t))^2-|v_{2}(t,t)|^2} , \\
& {\cal N}(t)=\frac{e^{-(1-\delta(t)|\alpha(t)|^2-\frac{1}{2}[\beta(t)(\alpha^*(t))^2+\beta^*(t) (\alpha^2(t)]}}{\sqrt{(1+v_{1}(t,t))^2-|v_{2}(t,t)|^2}}
\end{align}
\end{subequations}
that are fully determined by system propagator $u(t,t_0)$ and normal and anomalous 
fluctuation correlation functions $v_1(t,t)$ and $v_2(t,t)$ which are induced by initial 
squeezing environment.

For finitely many modes, the total normal-mode frequencies $\{ \omega^n_i\}$ are discrete, the solution of $u(t,t_0)$ is an oscillation 
function containing all normal-mode frequencies, 
\begin{align}
u(t,t_0)= \sum_{i=0}^{N_\env} c_i e^{-i\omega^n_i(t-t_0)} ~,~{\rm and }~~  \sum_i c_i=1.  \label{oscu}
\end{align}
because $u(t_0,t_0)=1$. Thus, for any finite $N_\env$, the system keeps energy exchanges with environment all time, no dissipation exists. 
The analytical solution of the normal and anomalous fluctuation functions $v_1(t,t)$ and $v_2(t,t)$ are given by,
\begin{subequations}
\label{s_naff}
\begin{align}
   v_1(\tau,t) & =\! \int^\tau_{t_0}  \!\!d\tau_1 \int^t_{t_0}  \!\! d\tau_2  u(\tau,\tau_1)\widetilde{g}(\tau_1,\tau_2) u^\ast(t,\tau_2) ,     
     \label{v1s}  \\
   v_2(\tau,t) & = \!\! \int^\tau_{t_0}  \!\!d\tau_1 \int^t_{t_0}  \!\! d\tau_2  u(\tau,\tau_1)\overline{g}(\tau_1,\tau_2) u(t,\tau_2) ,
    \label{v2s} 
\end{align}
\end{subequations}
which must also be oscillating functions, where the normal and anomalous fluctuation kernels linking the system mode 
to the environment are given by
$ \widetilde g_1(\tau,\tau')\!=\! \sum_k|V_{k}|^2 n_k e^{-i\omega_k(\tau-\tau')}$ and
$ \widetilde g_2(\tau,\tau')\!=\!\sum_k|V_{k}|^2 m_k e^{-i\omega_k(\tau+\tau'-2t_0)}$, respectively.
Note that even there is no dissipation, these identities remains true as a consequence of unitarity
of the total dynamics. 

For squeezing parameter $r_k\neq0$, we have 
$0<\delta(t)<1$ so that $\rho_\sys^2\neq\rho_\sys$ \cite{Yang2025}. 
Because the total density operator remains pure, we can write the total state in terms of Schmidt decomposition in any 
orthogonal system basis $\ket{\psi_{\sys i}}$ and orthogonal environment basis $\ket{\phi_{\env i}}$ \cite{Neumann1955}:
\begin{align}
\ket{\Psi_{\rm tot}(t)}  =\sum_j  \sqrt{\lambda_j(t)}\ket{\psi_{\sys j}} \ket{\phi_{\env j}},
\end{align}
where $\lambda_i(t) = \bra{\psi_{\sys i}} \rho_\sys(t) \ket{\psi_{\sys i}}$ is the probability in system   
basis $\ket{\psi_{\sys i}}$ with the environment (e.g. the measuring apparatus) pointing to the state $\ket{\phi_{\env i}}$. 
Obviously, $\lambda_j(t)$ is a function of $\delta(t), \alpha(t)$ and $\beta(t)$ of Eq.~(\ref{eq:delta}) and is uniquely 
determined by propagator and correlation function of Eqs.~(\ref{oscu})-(\ref{s_naff}). Then 
\begin{align}
S_\sys(t)& = -\Tr_\sys [\rho_\sys(t)\ln \rho_\sys(t)] \notag \\
&= -\sum_j \lambda_j(t) \ln \lambda_j(t)  =S_\env(t)  >0
\end{align}
is exactly the bipartite entanglement entropy. 
Because $\delta(t), \alpha(t)$ and $\beta(t)$ are multi-periodic functions with periods $T$ proportional to the inverse 
of differences between normal-modes, the entropy cannot monotonically increase, it exhibits Poincar\'{e}-type recurrences.  
Thus, a pure finite-mode environment establishes the step ``reversible whole $\to$ 
subsystem probability'' given by Eq.~(\ref{eq:squezquasi_g}), 
but 
no irreversible thermodynamic limit occurs, and no time arrow can be defined at this point.

Furthermore, if all the squeezing parameters $r_k=0$ in the environment, then 
\begin{align}
\rho_\sys(t)= \ket{u(t,t_0)\alpha_\sys} \bra{ u(t,t_0)\alpha_\sys}.  \label{csss}
\end{align} 
which is a pure coherent state propagating by $u(t,t_0)$. 
Because Glauber coherent state corresponds to the most classical state occupying a phase space value $\Delta x \Delta p 
=\hbar/2$ \cite{Zhang1990}, it shows that linear interaction of Eq.~(\ref{eq:finitemode}) itself cannot produce entanglement and cannot have 
``reversible whole $\to$ subsystem probability'' from classical reversible mechanics.
In fact, the above solutions are valid for arbitrary system initial state. Taking the Glauber coherent state as the initial state, 
we build a bridge connecting classical mechanics and showing the difference between quantum and 
classical dynamics, as shown by Eqs.~(\ref{eq:squezquasi_g}) and (\ref{csss}).

Case (ii): $N_\sys=1, N_\env \to \infty$ (environment becomes a continua with a continuum spectrum). 
Replace the finite environment modes $\omega_k$ by a continuum spectrum, then 
$V_i \to V(\omega)$ in Eq.~(\ref{eq:finitemode}), and the completely separable initial pure state becomes
\begin{equation}
 \ket{\Psi_{\rm tot}(t_0)}
 =\ket{\alpha_\sys}\otimes\prod_\omega \ket{s(\omega)},
 \quad s(\omega)=r(\omega)e^{i\theta(\omega)},   \label{inisq}
\end{equation}
where every environmental mode $\omega$ is still pure and has deterministic normal and anomalous covariances, 
\begin{align}
n(\omega)= \sinh^2 r(\omega)~,~m(\omega) =\!-\! \frac{1}{2}e^{i\theta(\omega)}\sinh 2r(\omega).
\end{align}
The solution of  $\rho_\sys(t)$ is still given by Eq.~(\ref{eq:squezquasi_g}) and (\ref{eq:delta}), 
but the corresponding dissipative,  normal and anomalous fluctuation kernels in Eqs.~(\ref{eq:u}) and (\ref{s_naff}) become
$g(\tau,\tau') =  \!\int^\infty_0 \! \frac{d\omega}{2\pi}J(\omega)e^{-i\omega(\tau-\tau')}$, 
$ \widetilde g_1(\tau,\tau') = \!\!\int^\infty_0 \! \frac{d\omega}{2\pi}J(\omega)n(\omega)e^{-i\omega(\tau-\tau')} $,
$\widetilde g_2(\tau,\tau') = \!\!\int^\infty_0  \! \frac{d\omega}{2\pi}J(\omega)m(\omega)e^{-i\omega(\tau+\tau'-2t_0)}$,
where $J(\omega)\!=\! \varrho(\omega)|V(\omega)|^2$, and $\varrho(\omega)$ is the density of states of the
environment. 

Now, the general solution of propagator $u(t,t_0)$ solved from Eq.~(\ref{eq:u}) becomes a dissipative propagator \cite{Zhang2012}
\begin{align}
u(t,t_0)= \sum_\ell Z_\ell e^{-i\omega_\ell(t-t_0)}+ \!\! \int \!\! \frac{d\omega}{2\pi} {\cal D}(\omega) e^{i\omega(t-t_0)}.  \label{gsu}
\end{align}
with the damping (level-broadening ) spectrum ${\cal D}(\omega)\!=\!\frac{J(\omega)}{[\omega\!-
\!\omega_0\!-\!\Delta(\omega)]^2\!+\! J^2(\omega)/4}$ and $\Delta(\omega) 
\!=\! {\cal P}\int \frac{d\omega'}{2\pi} \frac{J(\omega')}{\omega-\omega'}$, where ${\cal P}$ denotes the Cauchy principal value.
If there is no localized bound state (i.e. $\omega\!-\!\omega_0\!-\!\Delta(\omega)\! \neq 0$ in the region $J(\omega)=0$)
then the first term in Eq.~(\ref{gsu}) vanishes, which is valid usually for a weak or intermediate system-environment 
coupling. Thus, $u(t,t_0)$ decays non-exponentially  
in general \cite{Zhang2012} and $u(t\!\to\!\infty,t_0)=0$. The discrete recurrence time in finite $N_\env$ diverges, 
the initial state dependence on $\alpha_\sys$ disappears and {\it the time arrow} emerges.
The system retains stationary normal and anomalous moments, 
\begin{subequations}
\begin{align}
&  \lim_{t \rightarrow \infty}\! \Tr_\sys [a^\dag a \rho_\sys(t)] \!= \! \lim_{t \rightarrow \infty} \! 
v_1(t,t) = \!\! \int \! \frac{d\omega}{2\pi} {\cal D}(\omega) n(\omega) \equiv \overline n, \\
&  \lim_{t \rightarrow \infty} \! \Tr_\sys [a a \rho_\sys(t)] \! = \! \lim_{t \rightarrow \infty} \!   
v_2(t,t) = \!\! \int \! \frac{d\omega}{2\pi} {\cal D}(\omega) m(\omega) \equiv \overline m, 
\end{align}
\end{subequations}
as the steady-state solution of the quantum Boltzmann equation, the generalized equilibrium fluctuation-dissipation theorem
naturally holds for both the normal and anomalous moments. 
These mean values (the average particle number and the average squeezing) can be directly measured experimentally.

Taking the steady-state limit, Eq.~(\ref{eq:delta}) is simply reduced to $\alpha(t)  \to 0$ , $\beta(t) \to  
\beta_\sys=\overline{m}/\nu$, $\delta(t) \to \delta_\sys=(\overline{n}+\overline{n}^2 -|\overline{m}|^2)/\nu$,
${\cal N}(t) \to N_\sys= 1/\sqrt{\nu}$, with $\nu=(1+ \overline{n})^2 - |\overline{m}|^2$.
Thus, the reduced density operator of Eq.~(\ref{eq:squezquasi_g}) becomes a generalized Gibbs state,
\begin{align}
\lim_{t \to \infty} \rho_\sys(t) &= N_\sys e^{\frac{1}{2}\beta_\sys a^{\dag2}} 
\Big[\sum_{n=0}^\infty \delta_\sys^n |n\rangle \langle n| \Big]e^{\frac{1}{2}\beta^*_\sys a^2} \notag \\
 &=\frac{1}{Z_\sys}\exp \big(\!-\!\beta_{\rm eff}  \widetilde{H}_\sys\big) = \rho_\g .
\label{eq:squeezedG}
\end{align}
Note that $\sum_{n=0}^\infty \delta_\sys^n |n\rangle \langle n| = e^{(\ln \delta_\sys) a^\dag a}$, the second equality 
in Eq.~(\ref{eq:squeezedG}) is 
obtained by rearranging the ordering of the exponential operator products using the faithful representation
of group theory \cite{Zhang1990,Yao2024}. Also, because Eq.~(\ref{eq:squeezedG}) is typically a Gibbs-type state, 
we can always introduce an effective temperature, $\beta_{\rm eff}=1/k_\bc T_{\rm eff}$
which can be experimentally determined, to express the stationary state in a
standard Gibbs state form.  The renormalized system Hamiltonian in Eq.~(\ref{eq:squeezedG}) is \cite{Yao2024, explain2}
\begin{equation}
 \widetilde{H}_\sys=\hbar \widetilde{\omega}_\sys a^\dagger a  +
\frac{1}{2} \hbar(\eta_\sys a^{\dagger2}+\eta_\sys^*a^2),   \label{renormH}
\end{equation}
and the renormalized frequency $\widetilde{\omega}_\sys$ and the effective pairing $\eta_\sys$ are given by   
\begin{subequations}
\begin{align}
&\beta_{\rm eff} \hbar \widetilde{\omega}_\sys = - \frac{\overline{n}+1/2}{s}\ln\frac{s - 1/2}{s + 1/2}, \\
&\beta_{\rm eff} \hbar \eta_\sys =  \frac{\overline{m}}{s}\ln\frac{s - 1/2}{s + 1/2} ,
\end{align}
\end{subequations}
where $s=\sqrt{(\overline{n}+1/2)^2-|\overline{m}|^2}$. The partition function $Z_\sys=e^{-\beta_{\rm eff}\hbar \widetilde{\omega}_\sys/2}/
\sqrt{\overline{n}+\overline{n}^2 -|\overline{m}|^2}$.

The above derivation shows that a pure squeezing state environment with continuous-spectrum and no isolated bound-state 
produce a stationary state that is independent of its initial state. It represents a genuine 
Gibbs state with a renormalized Hamiltonian, an effective temperature and naturally a maximum-entropy property. 
Thus, we can calculate directly 
all ``macroscopic" physical quantities, such as the entropy and the internal energy from Eq.~(\ref{eq:squeezedG}), even though 
the system itself contains only one degrees of freedom. It is easy to show that
\begin{align}
S= -k_\bc \!\Tr_\sys [\rho_\g \ln \rho_\g] = \frac{1}{T_{\rm eff}} E + k_\bc\ln Z_\sys.   
\end{align}
where $E= \Tr_\sys \! \big[\widetilde{H}_\sys \rho_\g\big]$ is defined as the internal energy of the system. 
This naturally leads to the well-known fundamental equation of thermodynamics \cite{Callen1985}, but here it is  
rigorously derived from the unitary (reversible) quantum dynamics of a closed composite with a pure initial state,
\begin{align}
 F=- k_\bc T_{\rm eff} \ln Z_\sys =  E - T_{\rm eff} S 
\end{align}
is the free energy that is familiar in thermodynamics. 
Note that this quantum thermodynamics is produced from the initial pure squeezed vacuum of the environment.
Again, if the squeezing $r(\omega)=0$, one can immediately
find that $\overline{n}=\overline{m}=0$, so that the steady-state of Eq.~(\ref{eq:squeezedG}) can only be reduced to
a vacuum state $\lim_{t\to\infty}\rho_\sys(t)=\ket{0}\!\bra{0}$, irreversibility and thermalization cannot occur. 
Thus, the irreversibility is fully induced by the non-classicality (squeezing) of environmental vacuum. 
In other words, reversible Newtonian mechanics cannot produce irreversibility in linear systems. Now, this solution can be 
extended to continuum macroscopic systems.

Case (iii): $N_\sys \to\infty$ and $N_\env \to \infty$, which is a pure-state model 
with pure continuum system and environment that forms a continuum ``universe". 
The environment continuum spectrum can be assumed to across all the energies, then the system 
modes become a continuum part inside the ``universe". The initial pure state can be written as~\cite{explain5}
\begin{equation}
 \ket{\Psi_{\rm tot}(t_0)}
 \!=\! \prod_{\omega_\sys} \ket{\alpha(\omega_\sys)}\otimes\prod_{\omega_\env} \ket{s(\omega_\env)},
~ s(\omega_\env)=r(\omega_\env)e^{i\theta(\omega_\env)}.   \label{inisq}
\end{equation}
where $\{ \omega_\sys \} \in \Omega_\sys$ is a continuum frequency domain with a finite energy. 
As long as the continuum system is a part of the universe 
with a finite energy, all the formula presented in situation (ii) remains valid, and the steady state of the continuum system  
is still given by a generalized
Gibbs-type state of Eq.~(\ref{eq:squeezedG}), with the renormalized Hamiltonian
\begin{align}
 \widetilde{H}_\sys= \!\! \int_{\Omega_\sys} \!\!d\omega & d\omega' \big\{  \hbar \widetilde{\omega}_\sys(\omega,\omega') a^\dag(\omega) a(\omega') \notag \\
  & + \frac{1}{2}\hbar [\eta_\sys(\omega,\omega') a^\dag(\omega)a^\dag(\omega')+ {\rm H.c.}]\big\},
\end{align}
which naturally weave complex quantum entanglement and many-body coherence between originally unrelated subsystem modes.
Thus, we reach the following theorem:

\textit{Theorem (Gibbs limit from a pure initial composite).}
Consider systems with quadratic Hamiltonian~\eqref{eq:Htot} that has a continuous spectrum. 
If the total initial state and every factor state are pure, a part of them is in 
non-classical (e.g.~squeezing) state, the system-environment couplings are not too strong to 
generate localized bound states, then for every pure system initial state, the energy exchange 
between the system and the rest of universe dynamically drives it toward
\begin{align}
 \lim_{t\to\infty}\rho_\sys(t)
 =\frac{1}{Z_\sys}\exp \big(\!-\!\beta_{\rm eff} \widetilde{H}_\sys\big),
\label{eq:proposition}
\end{align}
while $\rho_{\rm tot}(t)$ remains pure and the system entropy approaches the maximum in the steady-state limit.
The stable renormalized quadratic Hamiltonian $\widetilde{H}_\sys$ can be obtained and the effective 
temperature $\beta_{\rm eff}$ can be defined from the dissipative propagator $\bm u$ and the fluctuation correlation 
function $\bm v$, the system initial state dependence disappears. The generalized fluctuation-dissipation theorem 
always holds. Otherwise, strong system-environment couplings and
gapped spectral densities can produce bound localized states with localized frequencies located in gaps, 
and thermalization fails.

The above results tell that unitarity guarantees the total entropy $S_{\rm tot}(t)=0$, but the exact partial trace (an exact non-perturbative 
renormalization) causes the dynamics of any subsystem follows a statistical probability description. The emergence of statistical
probability arises from the violation of internal causality in the subsystem dynamics.
This violation of  internal causality stems from partial vacuum squeezing or any other non-classical correlation 
present in the initial state \cite{Yang2025}. It is this violation of internal causality to the subsystem dynamics that causes irreversibility 
emergence and leads to entanglement between the system and the rest of the whole.
For a finite-mode (discrete-spectra) systems and environments, information can flow back, making entropy 
decrease and increase with oscillations. In a continua, locally accessible information dissipate to infinitely continuously 
distributed modes, no finite-dimensional subsystem can rephase all of them. Consequently, a macroscopic 
arrow of time naturally emerges from the dynamical evolution of general non-unitary open systems. 
It is the product of the interplay between subsystem reduction (non-perturbative renormalization to subsystem) 
and the continuous spectrum limit. This constitutes the mechanism for the emergence of the arrow of time from our 
theory of open quantum systems.

\section{Discussion and conclusion}

Hilbert's sixth problem~\cite{Hilbert1902} is a research program rather than one isolated proposition. 
The present work addresses the core issue of dynamically deducing statistical probability and thermodynamics 
from deterministic principle of mechanics, rather than the mathematical derivation 
from Newtonian mechanics to continuum hydrodynamics \cite{explain3}.
Within the bosonic and fermionic quadratic open quantum systems studied here, the construction starts from a closed, pure, 
deterministic composite and introduces neither a random force, a projective measurement, 
nor an initial thermal ensemble. Statistical probability arise as a dynamical production of the mixture of the reduced density 
operator. The Gibbs probability distribution follows for a continuous environment spectrum and absence of localized bound 
states in the reduced evolution of the system, where the generalized fluctuation-dissipation theorem hold all the time as a natural 
consequence of unitary evolution  of the whole composite. Our theory shows that any finite closed system is not possible 
to thermalize. The same theory identifies precise obstructions: a gap-induced bound state can retain memory on the initial state. 
Furthermore, it is the non-classical nature embedded in the initial pure state that dynamically leads to a Gibbs-type state 
at the end of the system evolution, regardless of whether the nonclassical initial pure state is carried by the system, 
the environment, or both.

The core methodology of our theory relies on deducing dynamically, rather than pre-defining, two fundamental Green's functions,
the dissipative propagating function $\bm u$ and the fluctuation correlating function $\bm v$~\cite{Jin2010,Lei2012,HuangTopological2020,HuangBrownian2022}.  These two Green's functions emerge 
naturally as physical quantities and encapsulate the full non-Markovian memory effects in our theory.
They are indeed one-to-one corresponding to the Schwinger-Keldysh non-equilibrium Green’s 
functions in many-body systems \cite{Schwinger1961,Keldysh1965,Chou1985,Zhang1992}, the latter are defined {\it a priori} to 
circumvent the direct calculation of the density operator of many-body systems in non-equilibrium physics,
a task that is often exceedingly difficult and has indeed never been done in general.
Thus, within our rigorous evolutionary framework, out theory thoroughly guarantees 
the complete physical consistency of dynamic evolution, initial state correlation, and environment feedback.
The next phase of this lone journey is extending this exact master equation framework to the
realistic quantum field theory of QED and QCD. However, under nonlinear interactions, the memory 
kernels no longer derive from a simple two-point correlation function, but naturally contains 
higher-order many-body correlation functions.  In fact, the solutions of Eq.~(\ref{qBes}) and (\ref{transpcurrent}) are
valid for the general many-body interacting systems, and the perturbative and non-perturbative approaches, 
such as diagrammatic expansion and loop expansion, can be directly utilized. However, genuinely interacting for  
calculating the reduced density operator itself requires further investigations. I leave this problem in further publications.

Our derivation does not use a probabilistic interpretation of wavefunctions as an initial input. In contrast, we start from
the rigorous unitary dynamics of the total Hamiltonian and a product of initial pure states with initially squeezed environment 
vacuum (or any other non-classical state) to show how the reduced density operator dynamically evolve into a fully mixed state 
(Gibbs-type state). This shows how probability arises inside a closed composite and provides the mixed-state foundation 
when measurement apparatus is treated as an environment. It also shows the dynamical generated measurement 
probability for local observables in quantum mechanics. 
It is indeed a microscopic dynamical realization of the measurement process originally proposed 
by von Neumann but he did not practically realize it~\cite{Neumann1955VI}. In this paper, we explicitly show that 
when the measured system and measurement apparatus are treated as a closed composite which obey the unitary 
dynamical evolution, the measured system itself will evolve into a mixed state after the measurement is performed.
This reveals the physical origin of probability in quantum mechanics, without requiring a probabilistic interpretation 
of the wavefunction. In fact, a wavefunction obeying the Schrödinger equation is always pure which should not 
possess probabilistic characteristics, thereby aligning with the property that the von Neumann entropy of a pure state is zero.

Moreover, both the quadratic model class and the spectral continuum requirement stated in the paper should be no difficult to
implement in experiments. Consequently, our theory provides not only a rigorous derivation of deterministic quantum mechanics 
to thermostatistic mechanics for Hilbert's sixth problem~\cite{Hilbert1902}, but also an exact, testable dynamical solution 
of the emergence of probability for quantum measurement processes~\cite{Neumann1955VI}. 
In fact, measuring two-time correlations in many-body systems, specifically covariances in quantum optics, is not difficult. 
The advanced experimental technologies, such as precision measurement,
non-demolition measurement, and faster detection with attosecond spectroscopy are currently available tools 
for accurately testing our theoretical consequences. 
We leave this practical measurement problem to experimentalists.  In additional, the question regarding one 
outcome in a single run (the so-called wavefunction collapes) remains a possible operational layer of 
measurement distinct from the generation of the probability itself presented in this paper.  

Our open quantum system perspective may provide a common language for more distant problems. If space-time geometry is 
a dynamical genesis of thermodynamic/entanglement description of quantum-field degrees of freedom, one may 
seek exact dynamical reductions from an underlying unitary evolution of field theory to local Gibbs states, as we done 
in this paper. In fact, a graded Sp(4) algebraic structure embedded in our Gibbs state gives a hint on the emergence of
space-time geometry. Furthermore, if an artificial-intelligence 
system is modeled as an open system that continuously exchanges energy, information, and memory with its environment, 
master equations in terms of non-Markovian correlation functions may provide a physically interpretable fundamental
theory of memory, forgetting, and steady learning. These are all promising areas for further development in the 
view of open quantum systems. As we have seen, approaching a generalized Gibbs state is just a natural outcome of our theory, 
how to maintain the system dynamically active through the localized bound states as the activation function is another
direction that we are pursuing. 

In conclusion, we have established one dynamical chain:
reversibility $\to$ internal causality breaking $\to$ entanglement $\to$ irreversibility $\to$  probability $\to$ thermalization.
The closed system always obeys deterministic Schr\"odinger or von Neumann evolution. The non-classical initial state  
and the system--environment interaction together leads to internal causality breaking of the subsystem dynamics, 
generates the entanglement between subsystems. 
This dynamical chain turns initially pure subsystem into mixed states with definite statistical weights. 
When one subsystem is enlarged to infinitely many degrees of freedom 
with a continuous spectrum and no bound state, the dissipative propagator decays, recurrences disappear, irreversibility emerges
and the exact reduced density operator approaches a renormalized Gibbs state where the fluctuation correlations naturally 
obey the generalized fluctuation--dissipation theorem. 
Within these well-defined modeling frameworks, the results constitute a dynamical solution to Hilbert's Sixth 
Problem, elucidating how statistical probability and thermodynamics emerge from deterministic quantum mechanics.
These findings also reveal how probability arises naturally during the process of quantum measurement, 
rather than being an additional interpretation imposed upon the wavefunction, as is the case in the 
standard understanding of quantum mechanics.

\begin{acknowledgments}
The results of this paper is partly based on my talk (mainly including the unpublished work) presented in the Symposium on Historical 
and Future Perspective of Physics held in Shanghai from Aug.~17th to Aug. 18th, 2026, which was
organized by Prof.~Yu Shi and Prof.~Yong-Shi Wu, as well as two lectures on $\lceil$Probability Theory:
the Logic of Sciences$\rfloor$ Book Club coordinated by Prof. Maoyuan Zhou, held on Aug. 21 and 28, 2026. 
I would like to thank many members in the Club for the intuitive discussions. In particular, I would like to express 
my special gratitude to Prof.~Yong-Shi Wu for many insightful questions he raised during my lectures.  
I would also like to thank my many former students and postdoctoral fellows, including Jun-Hong An, Matisse Wei-Yuan Tu, 
Jinshuang Jin, Chan U Lei, Heng-Na Xiong, Ping-Yuan Lo, Pei-Yun Yang, Hon-Lam Lai, A.~Ali, Fei-Lei Xiong, Yu-Wei Huang, 
Wei-Ming Huang, Chuan-Zhe Yao, Shuang-Kai Yang, and Yu-Juan Sun, for their contributions to the development of 
our open quantum system theory. 
Without their hard work, this established theory would not have been possible.  I want to thank ChatGPT for providing me an 
initial draft based on my talk and lectures, even though it causes chaos and randomly combines completely unrelated concepts 
and contents together as usual so that I had to rewrite it word by word, but honestly speaking, it does save me a lots of time in 
writing this paper.

\end{acknowledgments}

\end{document}